\documentclass{svjour2}                    
\smartqed  
\usepackage{graphicx}
\usepackage{amssymb}
\usepackage{amsmath}
\usepackage{times}
\usepackage[]{eufrak}
\usepackage{color}

\begin{document}

\title{Particle Motion and Electromagnetic Fields of  Rotating Compact
Gravitating Objects with Gravitomagnetic Charge
} 

\titlerunning{Particle Motion and Electromagnetic Fields of  Compact
Objects with Gravitomagnetic Charge}        

\author{A.A.~Abdujabbarov \and B.J.~Ahmedov \and V.G.~ Kagramanova}

\authorrunning{Abdujabbarov, Ahmedov, Kagramanova} 

\institute{A.A.~Abdujabbarov \at
              Institute of Nuclear Physics and
    Ulugh Beg Astronomical Institute,    Astronomicheskaya 33,
    Tashkent 100052, Uzbekistan \\
              \email{abahmadjon@yahoo.com}           
           \and
           B.J.~Ahmedov \at
              Institute of Nuclear Physics and
    Ulugh Beg Astronomical Institute
    Astronomicheskaya 33,
    Tashkent 100052, Uzbekistan \\
    The Abdus Salam International Centre
for Theoretical Physics, 34014 Trieste, Italy\\
              \email{ahmedov@astrin.uzsci.net}
            \and
            V.G.~ Kagramanova \at
            Institute for Physics, Oldenburg University, 26111
Oldenburg, Germany\\
    \email{kavageo@theorie.physik.uni-oldenburg.de}}

\date{Received: date / Accepted: date}

\maketitle

\begin{abstract}
\noindent The exact solution for the electromagnetic field
occuring when the Kerr-Taub-NUT compact object is immersed (i) in
an originally uniform magnetic field aligned along the axis of
axial symmetry (ii) in dipolar magnetic field generated by current
loop has been investigated. Effective potential of motion of
charged test particle around Kerr-Taub-NUT gravitational source
immersed in magnetic field with different values of external
magnetic field and NUT parameter has been  also investigated. In
both cases presence of NUT parameter and magnetic field shifts
stable circular orbits in the direction of the central gravitating
object. Finally we find analytical solutions of Maxwell equations
in the external background spacetime of a slowly rotating
magnetized NUT star. The star is considered isolated and in
vacuum, with monopolar configuration model for the stellar
magnetic field.


\end{abstract}

\section{Introduction}
\label{intro}

The existence of strong electromagnetic fields is one of the most
important features of rotating neutron stars observed as pulsars.
 {Starting the pioneering paper of Deutsch \cite{d55} it was proved that
due to the rotation of highly magnetized star the electric field
is induced.} The general relativistic effect of dragging of
inertial frames is very important in pulsar
magnetosphere~\cite{bes90,mt90} and  {is considered to be} a
source of additional electric field of general relativistic origin
(see, for example, ~\cite{mt92,kk00,ram01,ram01b}).

It was first shown by Ginzburg and Ozernoy~\cite{go64} that an
elecrically neutral black hole can not have an intrinsic magnetic
field. However assuming that black hole can be inserted in the
external  {uniform} magnetic field created by nearby source as
neutron star or magnetar, Wald~\cite{w74} found
 {for the first time} the exact solution of the
vacuum Maxwell equations for an asymptotically uniform magnetic
field. Then the properties  of black holes immersed in external
magnetic field were extensively studied  {by different} authors
(see, for example, \cite{ao02,dd84,p75,cv75,aliev_n} and for more
references \cite{nf89}).

Despite the absence of observational evidence for the existence of
gravitomagnetic monopole, that is of exotic space-time, called NUT
space (Newman, Unti and Tamburino~\cite{nut63}) at present, it is
interesting to study the electromagnetic fields and particle
motion in NUT space with the aim to get new tool for studying new
important general relativistic effects which are associated with
nondiagonal components of the metric tensor and have no Newtonian
analogues.

Here we study the electromagnetic fields in the Kerr-Taub-NUT
spacetime and in the surrounding space-time of slowly rotating
magnetized relativistic NUT star. Our approach is based on the
reasonable assumption that the metric of spacetime is known {,}
i.e. neglecting the influence of the electromagnetic field on the
gravitational one and finding analytical solutions of Maxwell
equations on a given, fixed background ( {estimations of
contribution of electromagnetic field energy into total energy
momentum could be found e.g. in paper~\cite{ra04}}).

We also study motion of test particle around Kerr-Taub-NUT source,
which is immersed either (i) in an originally uniform magnetic
field aligned along the axis of axial symmetry or (ii) in dipolar
magnetic field generated by current loop. We use Hamilton-Jacobi
equation to find influence of  {both} NUT parameter and magnetic
field  {on} the effective potential of motion of test charged
particles. Moreover here we completely ignore the pathologies
(existence of singularities along the axis or periodicity of the
time coordinate to avoid them and spacetime regions containing
closed timelike curves) of the spacetime metric due to fact that
NUT parameter is considered as small one.

The outline of the paper is as follows. In Section~\ref{bh} we
calculate the electric and magnetic fields  {generated} in the
Kerr-Taub-NUT spacetime  {following the method} of construction of
the vacuum solution to Maxwell equations in axially-symmetric
stationary spacetime suggested by Wald~\cite{w74}.

In Sections~\ref{motion} and ~\ref{loop} we consider the
separation of variables in the Hamilton-Jacobi equation and derive
the effective potential for the motion of charged particles around
Kerr-Taub-NUT source in a uniform and dipolar magnetic field
 {ge\-nerated} by current loop. These results are
used to obtain the basic equations  {go\-verning} the region of
marginal stability of the circular orbits and their associated
energies and angular momenta.

In Section~\ref{star} we look for stationary solutions of Maxwell
equations when the stellar magnetic field has monopolar
configuration which allows to find exact  {ana\-lytical} solution.
Section \ref{conclusion} is devoted to analysis of obtained
solutions for electric and magnetic fields exterior to the
rotating compact objects with NUT parameter.

Throughout, we use a space-like signature $(-,+,+,+)$ and a system
of units in which $G = 1 = c$ (However, for those expressions with
an astrophysical application we have written the speed of light
explicitly.). Greek indices are taken to run from 0 to 3 and Latin
indices from 1 to 3; covariant derivatives are denoted with a
semi-colon and partial derivatives with a comma.

\section{Kerr-Taub-NUT Source in a Uniform Magnetic Field}
\label{bh}

We consider electromagnetic fields of compact astrophysical
objects in Kerr-NUT spacetime which in a spherical coordinate
system $(ct,r,\theta,\varphi)$ is described by the metric (see,
for example,~\cite{dt02,betal03})
\begin{eqnarray}
\label{kerr_nut}
ds^2=-\frac{1}{\Sigma}(\Delta-a^2\sin^2\theta)dt^2+\frac{2}{\Sigma}
[\Delta\chi-a(\Sigma+a\chi)\sin^2\theta]dtd\varphi
\nonumber\\
+\frac{1}{\Sigma}[(\Sigma+a\chi)^2\sin^2\theta-\chi^2\Delta]d\varphi^2
+\frac{\Sigma}{\Delta}dr^2+\Sigma d\theta^2 \ ,
\end{eqnarray}
where parameters $\Sigma, \Delta$ and $\chi$ are defined by
\begin{equation}
\Sigma=r^2+(l+a\cos\theta)^2\ ,\quad \Delta =r^2-2Mr-l^2+a^2\ ,
\quad \chi=a\sin^2\theta-2l\cos\theta\ ,
\end{equation}
$l$ is the gravitomagnetic monopole momentum, $a=J/M$ is the
specific angular momentum of metric source with total mass $M$.

Here we will exploit the existence in this spacetime of a timelike
Killing vector $\xi^\alpha_{(t)}$ and spacelike one
$\xi^\alpha_{(\varphi)}$ being responsible for stationarity and
axial symmetry of geometry, such that they  {satisfy the} Killing
equations
\begin{equation}
\label{ke} \xi_{\alpha ;\beta}+\xi_{\beta;\alpha}=0\ ,
\end{equation}
and  {consequently the} wave-like equations (in vacuum spacetime)
\begin{equation}
\Box{\xi^\alpha}=0\ ,
\end{equation}
which gives a right to write the solution of vacuum Maxwell
equations $\Box A^\mu$ for the vector potential $A_\mu$ of the
electromagnetic field in the Lorentz gauge in the simple form
\begin{equation}
\label{pots} A^\alpha=C_1 \xi^\alpha_{(t)}+C_2
\xi^\alpha_{(\varphi)}\ .
\end{equation}
The constant $C_2=B/2$, where gravitational source is immersed in
the uniform magnetic field $\textbf{B}$ being parallel to its axis
of rotation. The value of the remaining constant $C_1$ can be
easily calculated from the asymptotic properties of spacetime
(\ref{kerr_nut}) at the infinity.

 Indeed in order to find the remaining constant one can use the
 electrical neutrality of the source
\begin{eqnarray}
\label{flux}  && 4\pi Q=0= \frac{1}{2}\oint
F^{\alpha\beta}{_*dS}_{\alpha\beta}=\nonumber\\
 && C_1 \oint \Gamma^\alpha_{\beta\gamma}u_\alpha m^\beta
\xi^\gamma_{(t)}(uk)dS +\frac{B}{2}\oint
\Gamma^\alpha_{\beta\gamma}u_\alpha m^\beta
\xi^\gamma_{(\varphi)}(uk)dS
\end{eqnarray}
evaluating the value of the integral through the spherical surface
at the asymptotic infinity. Here the equality
$\xi_{\beta;\alpha}=-\xi_{\alpha;\beta}=
-\Gamma^\gamma_{\alpha\beta}\xi_\gamma$ following from the Killing
equation was used, and element of an arbitrary 2-surface
$dS^{\alpha\beta}$ is represented in the form~\cite{ar03}
\begin{equation}
\label{surface} dS^{\alpha\beta}= - u^\alpha\wedge m^\beta (uk)dS
+ \eta^{\alpha\beta\mu\nu}u_\mu n_\nu\sqrt{1+(uk)^2}dS\ ,
\end{equation}
and the following couples
\begin{eqnarray}
&& m_\alpha =\frac{\eta_{\lambda\alpha\mu\nu}u^\lambda n^\mu
k^\nu} {\sqrt{1+(uk)^2}}, \quad n_\alpha
=\frac{\eta_{\lambda\alpha\mu\nu} u^\lambda k^\mu
m^\nu}{\sqrt{1+(uk)^2}}, \quad \\ && k^\alpha=-(uk)u^\alpha+
\sqrt{1+(uk)^2}\eta^{\mu\alpha\rho\nu}u_\mu m_\rho n_\nu \nonumber
\end{eqnarray}
are established between the triple $\{{\mathbf k},{\mathbf
m},{\mathbf n}\}$ of vectors, $n^\alpha$ is normal to 2-surface,
space-like vector $m^\alpha$ belongs to the given 2-surface and is
orthogonal to the four-velocity of observer $u^\alpha$, a unit
spacelike four-vector $k^\alpha$ belongs to the surface and is
orthogonal to $m^\alpha$, $dS$ is invariant element of surface,
$\wedge$ denotes the wedge product, $_*$ is for the dual element,
$\eta_{\alpha\beta\gamma\delta}$ is the pseudo-tensorial
expression for the Levi-Civita symbol $\epsilon_{\alpha \beta
\gamma \delta}$ .

Then one can insert $u_0=-(1-M/r)$, $m^1=(1-M/r)$, and asymptotic
values for the Christoffel symbols $\Gamma^0_{10}=M/r^2$ and
$\Gamma^0_{13}=-3J\sin^2\theta/r^2 -l(1-2M/r)\cos\theta/r$ in the
flux expression (\ref{flux}) and get the value of constant
$C_1=aB$. Parameter $l$ does not appear in constant $C_1$ because
the integral $\int_0^{\pi}\cos\theta\sin\theta d\theta =0$
vanishes.

Finally the 4-vector potential $A_\alpha$ of the electromagnetic
field will take a form
\begin{eqnarray}
\label{potential1}
A_0=-\frac{B}{\Sigma}\left\{\Delta\left(a-\frac{\chi}{2}\right)+
a\left[\frac{1}{2}(\Sigma+a\chi)- a^2\right]\sin^2\theta
\right\}=-\frac{BK}{\Sigma}\ ,\\
\label{potential2}
A_3=\frac{B}{\Sigma}\left\{\Delta\chi\left(a-\frac{\chi}{2}\right)+
 (\Sigma+a\chi)\left[\frac{1}{2}(\Sigma+a\chi)-
a^2\right]\sin^2\theta\right\}=\frac{BL}{\Sigma}\ .
\end{eqnarray}

Using solution for vector potential (\ref{pots}) in spacetime
(\ref{kerr_nut}) one could easily find expression for change of
the electrostatic energy of the charged particle, which is lower
to the horizon of gravitational source:
\begin{equation}
\epsilon=eA^\mu(\xi_{\mu(t)}+\Omega|_{hor}\xi_{\mu(\varphi)})
|_{hor}-eA^\mu\xi_{\mu(t)}|_{inf}=elB-eaB\ .
\end{equation}
At the horizon new timelike Killing vector \cite{carter,aliev_n}
\begin{eqnarray}
&&\psi_{\mu}|_{hor}=\xi_{\mu(t)}+\Omega|_{hor}\xi_{\mu(\varphi)}\ ,\nonumber\\
&&\Omega|_{hor}=\frac{a}{2Mr_+}\ ,\qquad r_+=M+\sqrt{M^2-a^2+l^2}\
\end{eqnarray}
is introduced since the timelike Killing vector $\xi$ becomes
spacelike inside ergoregion defined by $g_{00}=0$.

Upper limit for electric charge
\begin{equation}
Q=2aMB-2lMB\
\end{equation}
accreted by gravitational source will include in addition to the
contribution from Faraday induction effect arising from angular
momentum $a$ (see e.g. \cite{w74},\cite{aliev77}) the new term
being proportional to additional rotation coming from  NUT
parameter $l$. It is interesting to note that each parameter will
accrete the charges of opposite sign.

The orthonormal components of the electromagnetic fields measured
by zero angular momentum observers (ZAMO) with the four-velocity
components
\begin{eqnarray}
\label{uzamos} &&(u^{\alpha})_{_{\rm ZAMO}}\equiv\nonumber \\
&&
    \bigg(\sqrt{\frac{\left(\Sigma+a\chi\right)^2\sin^2\theta-\chi^2\Delta}
    {\Delta\Sigma\sin^2\theta}},0,0,
    {-\frac{\Delta\chi-a\left(\Sigma+a\chi\right)\sin^2\theta}
    {\sqrt{\Delta\Sigma\left[\left(\Sigma+a\chi\right)^2\sin^2\theta-
    \chi^2\Delta\right]}\sin\theta}}\bigg) \ ;
    \nonumber\\
&&(u_{\alpha})_{_{\rm ZAMO}}\equiv
    \bigg(- \sqrt{\frac{\Delta\Sigma\sin^2\theta}
    {\left(\Sigma+a\chi\right)^2\sin^2\theta-\chi^2\Delta}},0,0,0 \bigg) \
\end{eqnarray}
are given by expressions
\begin{eqnarray}
\label{e1} && E^{\hat r} =
-\frac{{2rB}\sin\theta}{\Sigma^2\sqrt{
\left(\Sigma+a\chi\right)^2\sin^2\theta-
\chi^2\Delta}}
\left\{\left[\Delta-\left(1-\frac{M}{r}\right)\Sigma-
a^2\sin^2\theta\right]\right.\nonumber\\
&&\hspace{1cm}\left.\times\left(\Sigma+a\chi\right)
\left(a-\frac{\chi}{2}\right)-
\frac{\Sigma}{2}\left[\chi\Delta-a(\Sigma+a\chi)\sin^2
\theta\right]\right\}\ ,\\
\label{e2} && E^{\hat \theta}=
\frac{B\sin^2\theta}{\Sigma^2\sqrt{\Delta(\left(\Sigma+a
\chi\right)^2\sin^2\theta- \chi^2\Delta})} \nonumber \\
&& \hspace{1cm}\times\bigg[\left\{\Delta(l+a\cos\theta)
+2a(\Sigma+a\chi-2a^2)
\cos\theta\right\}\Sigma(\Sigma+a\chi)\nonumber\\
&&\hspace{1cm}-\left.\left\{\Sigma(
\Sigma+a\chi-2a^2)-2K(l+a\cos\theta)\right\}\frac{\chi
\Delta}{\sin^2\theta}\right] ,\\
\label{m1} && B^{\hat r} =
\frac{B\sin\theta}{\Sigma\sqrt{\left(\Sigma+a\chi
\right)^2\sin^2\theta-\chi^2\Delta}}  \bigg[\chi\Delta(l+
a\cos\theta)\nonumber\\
&&\hspace{1cm}\left.-(\Sigma-a\chi)(\Sigma+a\chi-2a^2)
\cos\theta-\frac{2K}{\Sigma}
(\Sigma+a\chi)(l+a\cos\theta)\right] ,\\
\label{m2} && B^{\hat\theta} = -
\frac{2rB\Delta}{\Sigma^2\sqrt{\Delta(\left(\Sigma+
a\chi\right)^2\sin^2\theta-\chi^2\Delta})}
\times\left\{\left[\Delta-\left(1- \frac{M}{r}\right)\Sigma
\right.\right.
\nonumber\\
&&\hspace{1cm}\left.-a^2\sin^2\theta\bigg]
\left(a-\frac{\chi}{2}\right)\chi-\frac12\Sigma^2
\sin^2\theta\right\}\ ,
\end{eqnarray}
which depend on angular momentum and NUT parameter in complex way.
Astrophysically it is interesting to know the limiting cases of
expressions (\ref{e1})--(\ref{m2}), for example in either linear
or quadratic approximation ${\cal O}(a^2, l^2/r^2)$ in order to
give physical interpretation  of possible physical processes near
the slowly rotating relativistic compact stars, where they take
the following form:
\begin{eqnarray}
&& E^{\hat r} =
\frac{B}{r}\left\{(2l+a\cos\theta)\cos\theta-\frac{M
(12l\cos\theta+a(1+3\cos 2\theta))}{2 r}\right\} , \\
&& E^{\hat \theta}=
\frac{B\sin\theta}{r}\left\{l\left(1+\frac{2\cos\theta}
{\sin^2\theta}\right)+a(3\cos\theta-1)\right\} , \\
&& B^{\hat r} =-B\cos\theta\bigg\{1-\frac{1}{2r^2}
\bigg(-4(l^2-a^2+al\cos\theta)
 \nonumber\\
&&\hspace{1cm}
-4l^2\cot^2\theta+a(2l+3a\cos\theta)\sin\theta\tan\theta\bigg)\bigg\}
 ,\\
&& B^{\hat\theta} = B\sin\theta\bigg\{1-\frac{M}{r}+ \frac{1}{16
r^2\sin^2\theta}(-a^2+4l^2-4M^2\nonumber\\
&&\hspace{1cm}-8al\cos\theta+4(7l^2+M^2)\cos 2\theta +8al\cos
3\theta+a^2\cos 4 \theta)\bigg\} \ .
\end{eqnarray}

In the limit of flat spacetime, i.e. for $M/r\rightarrow 0$,
$Ma/r^2\rightarrow 0$ and $l^2/r^2\rightarrow 0$, expressions
(\ref{e1})--(\ref{m2}) give
\begin{eqnarray}
\label{limit_B_1} && \lim_{M/r, Ma/r^2, l^2/r^2\rightarrow 0}
B^{\hat r}=-B\cos\theta
    \ ,
 \lim_{M/r, Ma/r^2, l^2/r^2\rightarrow 0}
B^{\hat\theta}=B\sin\theta
     \ ,
\\ \nonumber \\
\label{limit_E} && \lim_{M/r, Ma/r^2, l^2/r^2\rightarrow 0}
E^{\hat r}=\lim_{M/r, Ma/r^2, l^2/r^2\rightarrow
0}E^{\hat\theta}=0 \ .
\end{eqnarray}
 As expected, expressions
(\ref{limit_B_1})--(\ref{limit_E}) coincide with the solutions for
the homogeneous magnetic field in Newtonian spacetime.

Finally we would like to show that two form of electromagnetic
field tensor will take more simplified form:
\begin{eqnarray}
\label{F}
&&F=\frac{B}{\Sigma^2}\left[2Kr-\Sigma(r-M)(2a-\chi)\right]
\omega^1 \wedge
\omega^0+\frac{B(l+a\cos\theta)\sin\theta}{\Sigma^2\Delta^{1/2}}
\nonumber\\
&&\qquad\left[ \Delta(\Sigma+2a^2-a\chi)+
a^2(\Sigma+a\chi-2a^2)\sin^2
\theta-2aK \right]\omega^2 \wedge \omega^0\nonumber\\
&&\qquad-\frac{Br\Delta^{1/2}\sin\theta}{\Sigma}\omega^1 \wedge
\omega^3+\frac{B\cos\theta}{\Sigma^2}\left[-a\chi(\Sigma+a\chi-2a^2)
\cos\theta+\right.\nonumber\\
&&\qquad\left.(l+a\cos\theta)[a(\Sigma+a\chi-2a^2)\sin^2
\theta+2\Delta a-\Delta\chi]\right]\omega^2\wedge\omega^3
\end{eqnarray}
in the orthonormal Carter-type frame \cite{aliev77}:
\begin{eqnarray}
&&\omega^0=\left(\frac{\Delta}{\Sigma}\right)^{1/2} (dt-\chi
d\varphi)\ ,\qquad \omega^1=\left(\frac{\Sigma}{\Delta}\right) dr
\ ,
\nonumber\\
&& \omega^2=\Sigma^{1/2}d\theta\ ,\qquad \omega^3=\frac{\sin
\theta}{\Sigma^{1/2}}\left[adt-(r^2+a^2+l^2)d\varphi\right]\ .
\end{eqnarray}

In the limiting case when NUT parameter $l\rightarrow0$ the
components of field tensor (\ref{F}) almost coincide with the
expressions (3.9) of paper \cite{w74}. In the limit of flat
spacetime, expressions for components of electromagnetic field
derived using tensor (\ref{F}) take Newtonian limits
 (\ref{limit_B_1})--(\ref{limit_E}).

\section{Motion of charged particles}\label{motion}

It is very important to investigate in detail the motion of
charged particles around a rotating compact source with NUT
parameter in an external magnetic field given by 4-vector
potential (\ref{potential1}) and (\ref{potential2}) with the aim
to find astrophysical evidence for the existence of
gravitomagnetic charge.

The Hamilton-Jacobi equation
\begin{equation}
\label{Ham-Jac-eq} g^{\mu\nu}\left(\frac{\partial S}{\partial
x^\mu}+eA_\mu\right)\left(\frac{\partial S}{\partial
x^\nu}+eA_\nu\right)=m^2\ ,
\end{equation}
for motion of the charged test particles with mass $m$ and charge
$e$ is applicable as a useful computational tool only when
separation of variables can be effected.

Since Kerr-Taub-NUT spacetime admits such separation of variables
(see e.g. \cite{dt02}) we shall study the motion  around source
described with metric (\ref{kerr_nut})  using the Hamilton-Jacobi
equation when the action $S$ can be decomposed in the form
\begin{equation}
\label{action}
S=-{\cal E}t+{\cal L}\varphi+S_{\rm
r\theta}(r,\theta)\ ,
\end{equation}
since the energy $\cal E$ and the angular momentum $\cal L$ of a
test particle are constants of motion in spacetime
(\ref{kerr_nut}). This is an extension of approach developed in
the paper \cite{ao02} to the case with nonvanishing NUT parameter.

Therefore the Hamilton-Jacobi equation (\ref{Ham-Jac-eq}) with
action (\ref{action}) implies the equation for inseparable part of
the action:
\begin{eqnarray}
&&\Delta\left(\frac{\partial S_{\rm r\theta}}{\partial
r}\right)^2+\left(\frac{\partial S_{\rm r\theta}}{\partial
\theta}\right)^2-\frac{(\Sigma+a\chi)^2\sin^2\theta-\chi^2\Delta}{
\Delta\sin^2\theta}\left({\cal E}+\frac{eBK}{\Sigma}\right)^2 \nonumber\\
&&\hspace{1.4cm}
-\frac{2(\Delta\chi-a(\Sigma+a\chi)\sin^2)}{\Delta\sin^2\theta}
\left({\cal E}+\frac{eBK}{\Sigma}\right)\left({\cal L}+\frac{eBL}{\Sigma}\right) \nonumber\\
&&\hspace{1.4cm}
+\frac{\Delta-a^2\sin^2\theta}{\Delta\sin^2\theta}\left({\cal L}+
\frac{eBL}{\Sigma}\right)-m^2\Sigma=0\ .
\end{eqnarray}

It is not possible to separate variables in this equation in
general case but it can be done for the motion in  the equatorial
plane $\theta = \pi/2$ when the equation for radial motion takes
form
\begin{equation}
\left(\frac{dr}{d\sigma}\right)^2={\cal E}^2-1-2V({\cal E},{\cal
L},r,b,a,l)\ .
\end{equation}

Here $\sigma$ is the proper time along the trajectory of a
particle, ${\cal E}$ and ${\cal L}$ are energy and angular
momentum per unit mass $m$ and
\begin{eqnarray}
\label{eff_potential}
&& V({\cal E},{\cal
L},r,b,a,l)=\nonumber\\
&&\hspace{1.4cm} -\left({\cal E}+\frac{b K}{M\Sigma}\right)\frac{b
K}{M\Sigma}+\frac{b^2
K^2}{2M^2\Sigma^2}-\frac{a^2}{\Sigma}\left(1+\frac{a^2}{2\Sigma}
\right)\left({\cal E}+\frac{b
K}{M\Sigma}\right)^2\nonumber\\
&&\hspace{1.4cm} -\frac{2l^2+2Mr-a^2}{2\Sigma}+
\frac{\Delta-\Sigma-a^2}{\Sigma^2}a\left({\cal E}+\frac{b
K}{M\Sigma}\right)\left({\cal L}+\frac{b
L}{M\Sigma}\right)\nonumber\\
&&\hspace{1.4cm} -\frac{\Delta-a^2}{2\Sigma^2}\left({\cal
L}+\frac{b L}{M\Sigma}\right)^2
\end{eqnarray}
can be thought of as an effective potential of the radial motion
which depends on additional dimensionless parameter
\begin{equation}
b=\frac{eBM}{m}
\end{equation}
being responsible for the relative influence of a uniform magnetic
field on the motion of the charged particles which maybe valuable
even for small values of the magnetic field strength \cite{ao02}.

Figure 1 shows radial dependence of effective potential
(\ref{eff_potential}) for different values of NUT parameter
$\tilde{l}=l/M$. From this dependence one can obtain modification
of radial motion of charged particle in the equatorial plane in
the presence of NUT parameter. As it is seen from the figure the
gravitomagnetic monopole momentum changes the shape of effective
potentials  when external magnetic field is not strong (Fig.1, a).
In the case of strong external magnetic field (Fig.1, b),
influence of gravitomagnetic monopole momentum is negligible.

{Figure 2 shows radial dependence of effective potential
(\ref{eff_potential}) for different values of $b$ for fixed value
of NUT parameter $\tilde{l}=0.5$.  Motion of charged particle in
the presence of this kind of effective potential can be explained
as follows: in the presence of external magnetic field in addition
to stable circular orbits unstable circular orbits could appear
due to the appearance of maximum in the graphs of effective
potential.}  From the potential we can infer the qualitative
structure of the particles orbits. As it is seen from the figure
the potential carries the repulsive character. It means that the
particle coming from infinity and passing by the source will not
be captured: it will be reflected and will go to infinity again.
For small values of electromagnetic field particles can follow
both bound and unbound orbits depending on their energy. As
external electromagnetic field increases interesting feature
arises: the orbits start to be only parabolic or hyperbolic and no
more circular or elliptical orbits
exist.

\begin{figure*}
  \includegraphics[width=0.75\textwidth]{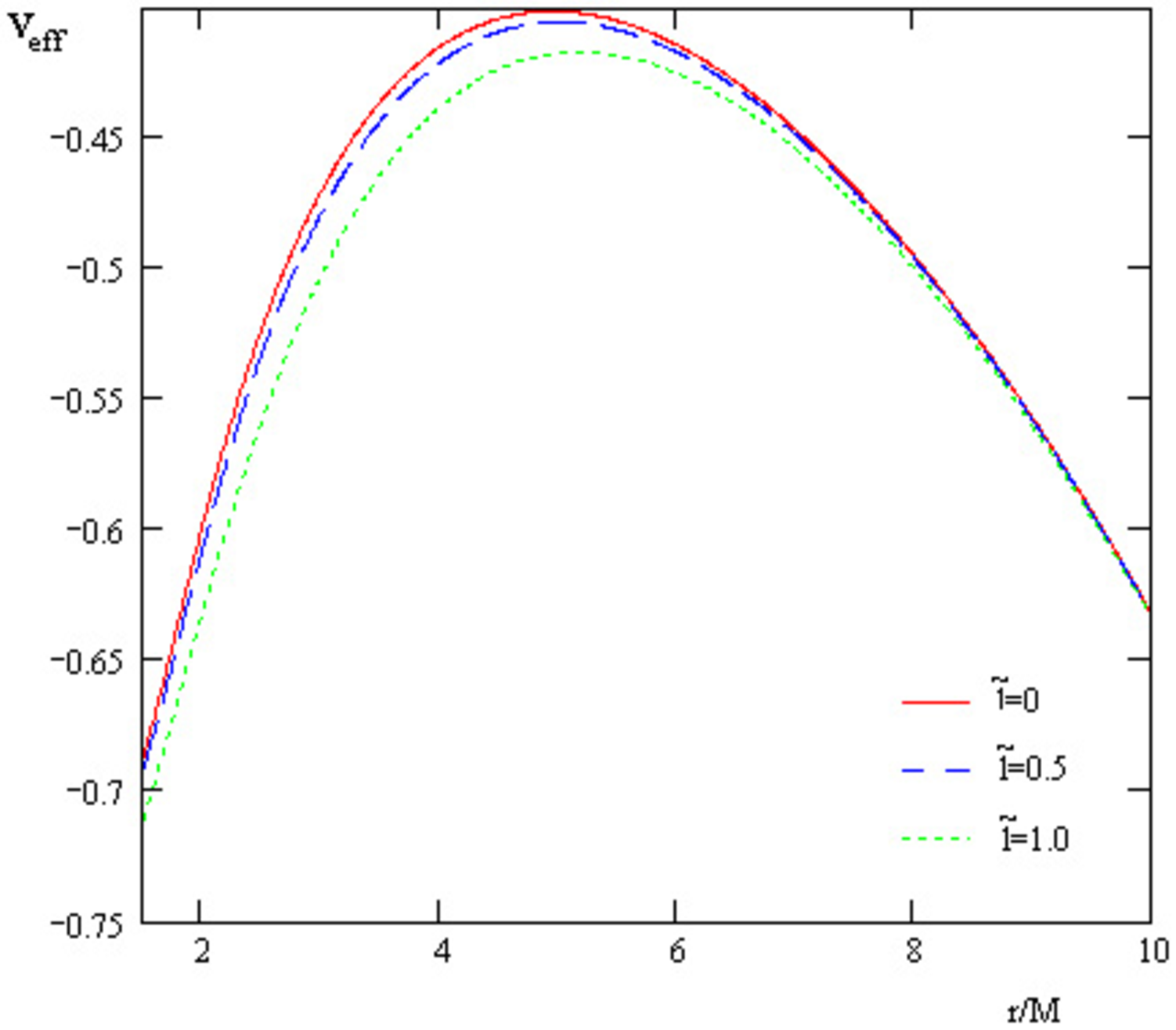}
   a)
  \includegraphics[width=0.75\textwidth]{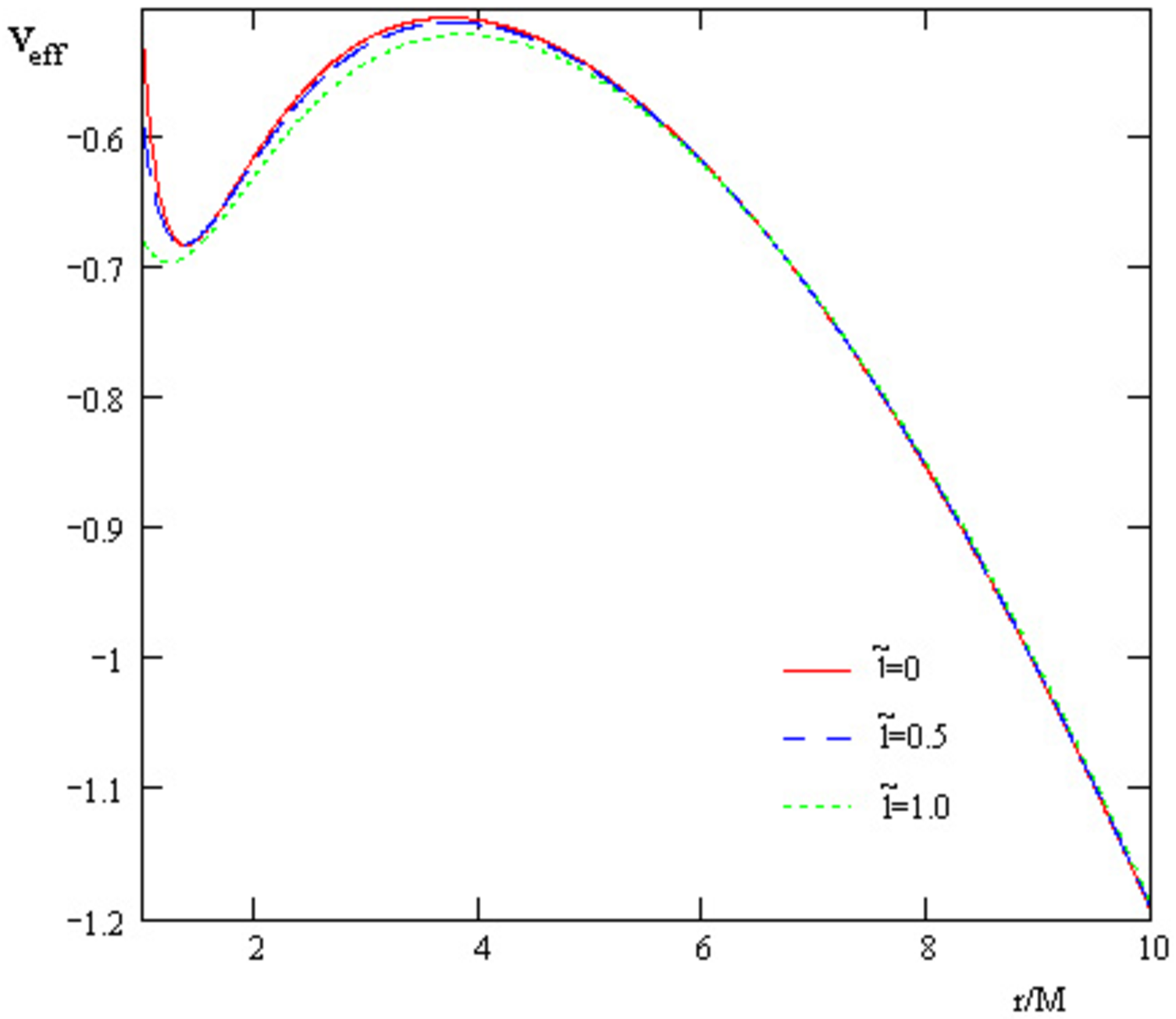}
   b)

\caption{{Radial dependence of the effective potential of radial
motion of charged particle near the Kerr-Taub-NUT source immersed
in external uniform magnetic field for different values of NUT
parameter $\tilde{l}=l/M$ for two cases of magnetic parameter: a)
$b=0.1$ and b) $b=0.15$.}}
\label{fig:1}       
\end{figure*}

\begin{figure}
  \includegraphics[width=0.75\textwidth]{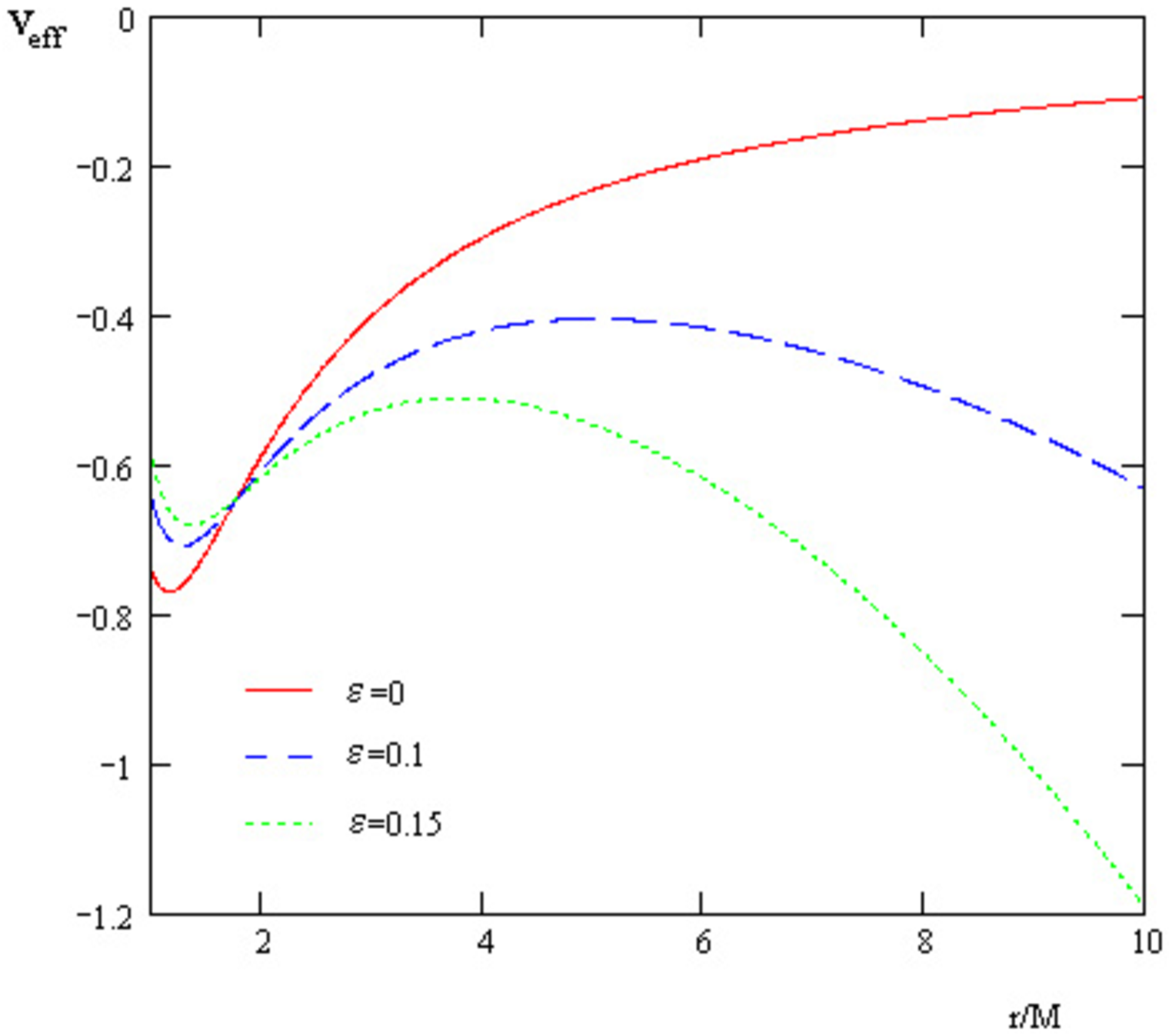}

\caption{{Radial dependence of the effective potential of radial
motion of charged particle near the Kerr-Taub-NUT source immersed
in external uniform magnetic field for the different values of
magnetic field in the  case when NUT parameter $\tilde{l}=0.5$}}
\label{fig:2}       
\end{figure}

\subsection{{Marginally stable circular orbits}}\label{stabcirc}

Special interest for the accretion theory of test particles around
a rotating compact source with NUT parameter in a magnetic field
is related to study of circular orbits  which are possible in the
equatorial plane $\theta=\pi/2$ when $dr/d\sigma$ is zero. It is
well known that in the presence of NUT parameter trajectory of
particle lies in the cone with $\cos\theta=\pm|2l{\cal E}/{\cal
L}|$ (see e.g. \cite{betal03}). In the case of energy and momentum
of the particle are equal to ${\cal E}=0.9\,\ {\cal L}=4.3$
respectively, and NUT parameter  $l\approx 0.5$, then we get
$90^\circ-10^\circ\leq\theta\leq90^\circ+10^\circ$, which allows
us to neglect deviation and consider the motion in the almost
equatorial plane. Consequently the right hand side of equation
(\ref{eff_potential}) vanishes:
\begin{equation}
\label{1steq_sys} {\cal E}^2-1-2V({\cal E},{\cal L},r,b,a,l) = 0
\end{equation}
along with its first derivative with respect to $r$
\begin{equation}\label{2ndeq_sys}
\frac{\partial V({\cal E},{\cal L},r,b,a,l)}{\partial r} = 0\ .
\end{equation}
%
In papers ~\cite{pv78} and ~\cite{p80} the authors have appealed
to a numerical analysis of analogues of equations
(\ref{1steq_sys}) and (\ref{2ndeq_sys}), when NUT parameter is
equal to zero, for different values of the constants of motion,
orbital radii, the rotation parameter $a$, as well as the strength
parameter of the magnetic field. {The problem of existence of
stable orbits in NUT space--time~\cite{manko} and other properties
of particles motion are also discussed in our preceding
paper~\cite{lebo}.}


The radius of marginal stability, the associated energy and
angular momentum of the circular orbits can be derived from the
simultaneous solution of the condition
\begin{equation}
\label{3rdeq_sys} \frac{\partial^2 V({\cal E},{\cal
L},r,b,a,l)}{\partial r^2}=0\
\end{equation}
and equation (\ref{1steq_sys}).

Combining equations (\ref{1steq_sys}) and (\ref{2ndeq_sys}) one
could find the energy
\begin{equation}\label{soltnE2}
{\cal E}=-\frac{ab}{M}+\gamma\pm\sqrt{\lambda}\ ,
\end{equation}
and the angular momentum
\begin{equation}\label{soltnL2}
{\cal L}=-\frac{b}{2M}(l^2+a^2)+\eta\pm(4ab-12r
\gamma)\sqrt{\lambda}\
\end{equation}
of a test particle, where the following notations
\begin{equation}
\gamma=\frac{2ab(r-\beta)}{3r^2-l^2-6\beta r} \quad, \qquad
\beta=r\frac{6r-4M}{12r-4M}\ ,
\end{equation}
\begin{eqnarray}
&& \lambda=[3r^2-l^2-6\beta
r]^{-1}\left[\frac{5b^2r^3}{4M^2}(r-4\beta)-
\frac{2b^2r^2}{M}(r-3\beta)\right.\qquad\qquad\qquad\qquad
\nonumber\\
&&\hspace{1.4cm} +3r\left(1-\frac{3b^2l^2}{4M^2}\right)
(r-2\beta) \nonumber\\
&&\hspace{1.4cm} \left.+2\left(\frac{l^2-a^2}{M}
b^2-2M\right)(r-\beta)+a^2-2l^2\right]
\end{eqnarray}
and
\begin{eqnarray}
\eta=4ab\gamma-6r\lambda-6r\gamma^2-
\frac{5b^2r^3}{M^2}+\frac{6b^2r^2}{M}
\qquad\qquad\qquad\qquad\nonumber\\
-6r\left(1-\frac{3b^2l^2}{4M^2}\right)-
2\left(\frac{l^2-a^2}{M}b^2-2M\right)\
\end{eqnarray}
were introduced.

Inserting now (\ref{soltnE2}) and (\ref{soltnL2}) into equation
(\ref{3rdeq_sys}) one could obtain the basic equation
\begin{eqnarray}
&& (r^3-l^2r+2a^2M)(\gamma\pm\sqrt{\lambda})^2+2a
b(2l^2-r^2)(\gamma\pm\sqrt{\lambda})
\nonumber\\
&&\hspace{1.4cm} +4aM(\gamma\pm\sqrt{\lambda})\nonumber\\
&&\hspace{1.4cm} \times(\eta\pm(4a b-12r\gamma)\sqrt{\lambda})+2b
r^2\left(\frac{r}{M}-1\right)
(\eta\pm(4ab-12r\gamma)\sqrt{\lambda})\nonumber\\
&&\hspace{1.4cm} +\frac{b^2r^5}{4M^2}-\frac{b^2r^4}{2M}+
\left(1-\frac{3b^2l^2}{4M^2}\right)r^3+
\left(\frac{l^2-a^2}{M}b^2-2M\right)r^2\nonumber\\
&&\hspace{1.4cm} +(a^2-2l^2)r+2Ml^2-\frac{4a^2l^2b^2}{M}=0\ .
\end{eqnarray}

The numerical solutions of this equation will determine the radii
of stable circular orbits for non-rotating NUT source immersed in
uniform magnetic field as functions of the NUT parameter $l$, the
angular momentum $a$, as well as of the influence parameter of the
magnetic field $b$. {In the Table 1 it is shown list of numerical
solutions for radii of stable circular orbits of particles for
different values of NUT parameter and external magnetic field.
With the increase of the gravitomagnetic monopole momentum, radii
of stable circular orbits shifts to gravitational object, while
external field also displace orbits to gravitational source}
\begin{table}\label{unifmag}
\caption{{Marginally circular orbits around non-rotating NUT
source immersed in uniform magnetic field} }
\begin{tabular}{llllll}
\hline\noalign{\smallskip}
$\tilde{l}$ & 0 & 0.01 &  0.1 &  0.3 &  0.5   \\
\noalign{\smallskip}\hline\noalign{\smallskip}
$b=0.1$ & 3.66650 & 3.66460 & 3.66154 & 3.62072 & 3.53294 \\
$b=0.2$ & 2.83292 & 2.83288 & 2.82876 & 2.79448 & 2.72020 \\
$b=0.3$ & 2.55504 & 2.55500 & 2.55162 & 2.52360 & 2.46338 \\
$b=0.4$ & 2.41614 & 2.41610 & 2.41326 & 2.38984 & 2,33994 \\
$b=0.5$ & 2.33282 & 2.33280 & 2.33036 & 2.31032 & 2.26794 \\
\noalign{\smallskip}\hline
\end{tabular}
\end{table}

\section{Motion of Charged Particle in the Field of Current
Carrying Loop Located Near the Kerr-Taub-NUT Source}\label{loop}

Consider a motion of test particle in the electromagnetic field
created by toroidal currents of ionized matter rotating in
accretion discs~\cite{camenzind} in Kerr-Taub-NUT spacetime. We
shall employ internal solution of ~\cite{peterson74}. In
particular, of the full multipole solution for magnetic field
provided in ~\cite{peterson74}, we shall here focus our attention
on the dominant (dipolar) term, which is given by
\begin{equation}
\label{loop_pot}A_\alpha=-\frac38\delta^\varphi_\alpha\frac{\mu
r^2\sin^2\theta}{M^3}
\left[\ln\left(1-\frac{2M}{R}\right)+\frac{2M}{R}
\left(1+\frac{2M}{R}\right)\right],
\end{equation}
where
$$
\mu=\pi R^2\left(1-\frac{2M}{R}\right)^{1/2}I
$$
is the modulus of the dipole moment due to the current $I$ in the
loop and $R$ is the radius of the loop, which is considered to be
approximately equal to $6M$ (because of small values of $l$ and
$a$ it will be not essentially modified) (see ~\cite{peterson74}).

Using Hamilton-Jacobi equation (\ref{Ham-Jac-eq}) and potential
(\ref{loop_pot}) one can find expression for effective potential
for radial motion of charged particle in the equatorial plane
$(\theta=\pi/2)$. It takes the following form
\begin{equation}\label{efpotlorig}
V_{eff}=-\frac{a^2}{\zeta}\left(\frac12+\frac{\eta}{\zeta}
\right){\cal E}^2-\frac{2\eta-a^2}{2\zeta}+\frac{2a{\cal
E}\eta}{\zeta^2} \left({\cal L}-\frac{\varepsilon}{M} r^2
\right)+\frac{\zeta-2\eta}{2\zeta^2}\left({\cal
L}-\frac{\varepsilon}{M} r^2 \right)^2,
\end{equation}
where we use the notations
$$
\zeta=r^2+l^2,\ \ \eta=Mr+l^2,
$$
and
$$
\varepsilon=-\frac38\frac{e\mu}{mM^2}
\left[\ln\left(1-\frac{2M}{R}\right)+\frac{2M}{R}
\left(1+\frac{2M}{R}\right)\right] \ .
$$
In the case, when rotational parameter $a$ is very small effective
potential takes following form:
\begin{equation}\label{efpotl}
V_{eff}=-\frac{\eta}{\zeta}+\frac{\zeta-2\eta}{2\zeta^2}\left({\cal
L}-\frac{\varepsilon}{M} r^2 \right)^2,
\end{equation}
which coincides with effective potential of radial motion in the
equatorial plane in the Schwarzschild metric (see for example
~\cite{hartle}) if one puts $\zeta=r^2,\ \eta=Mr,\ $ and
$\varepsilon=0$ into equation (\ref{efpotl}).

{Plots in figure 3 show the radial dependence of effective
potential, which is governed by equation (\ref{efpotlorig}), for
the different values of $\tilde{l}$. Like in the case of uniform
external magnetic field, described in the previous section, in
this case gravitomagnetic monopole momentum has an influence onto
the effective potential, if magnetic field is weak (fig.3, a). If
magnetic field of  loop is strong influence of gravitomagnetic
monopole momentum is negligible (fig.3, b).}

{Consider now stable circular orbits of charged particles as it
was done in Section \ref{motion}. We shall repeat calculations,
which are done in Subsection \ref{stabcirc}, in this case. Using
expression (\ref{efpotl}) as effective potential we find numerical
solutions for radii of stable circular orbits of charged
particles. From results shown in the Table 2 and Table 3 (for
anti-Larmor and Larmor orbits, respectively) one can obtain that
with the increase of the gravitomagnetic monopole momentum radii
of anti-Larmor and Larmor orbits shift to loop while with the
increase of the electric current (creating dipolar magnetic field)
of loop the stable orbits shift to the gravitational object. These
results may be useful to determine NUT parameter from
astrophysical observations.}

\begin{figure*}
  \includegraphics[width=0.75\textwidth]{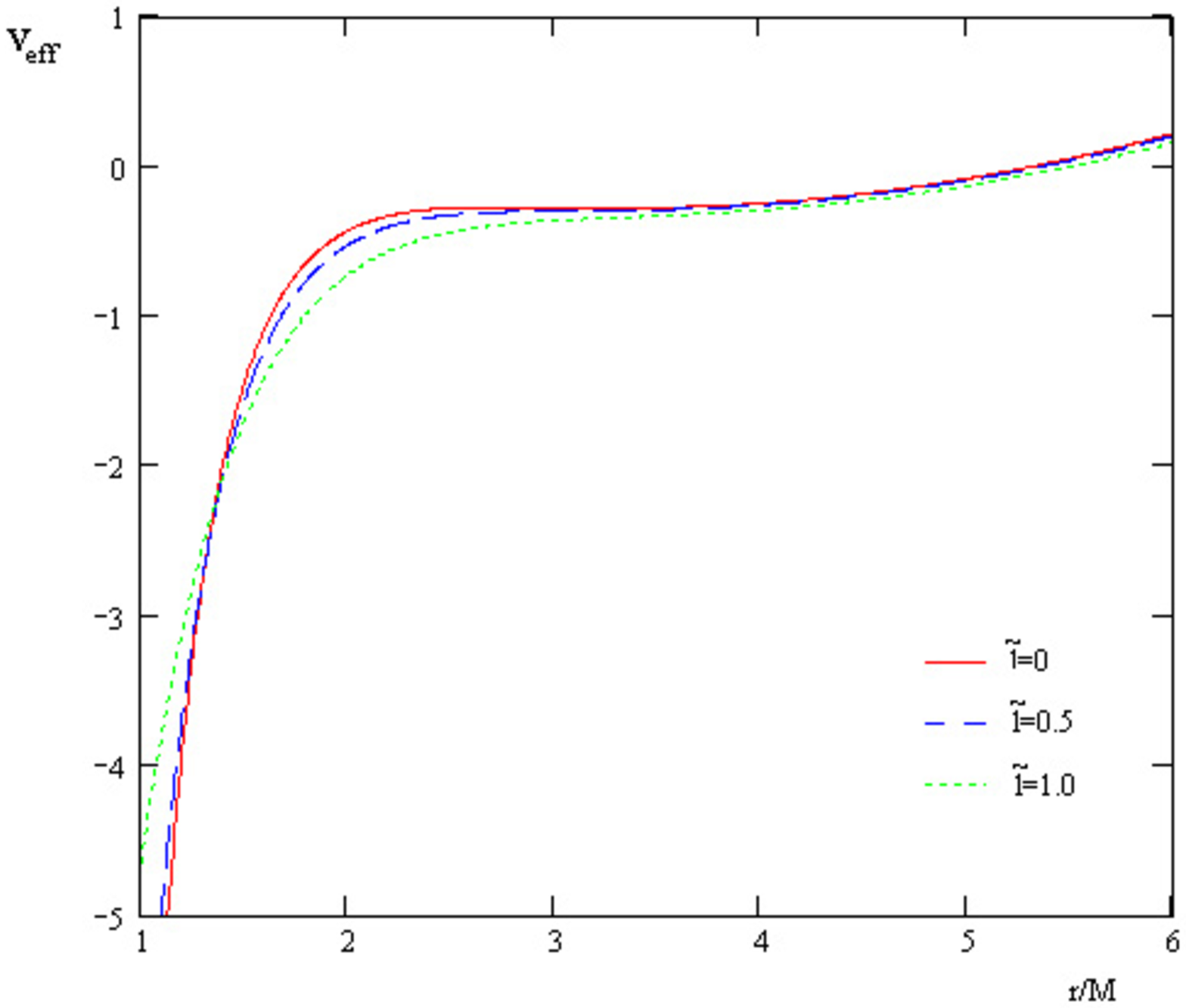}
  \includegraphics[width=0.75\textwidth]{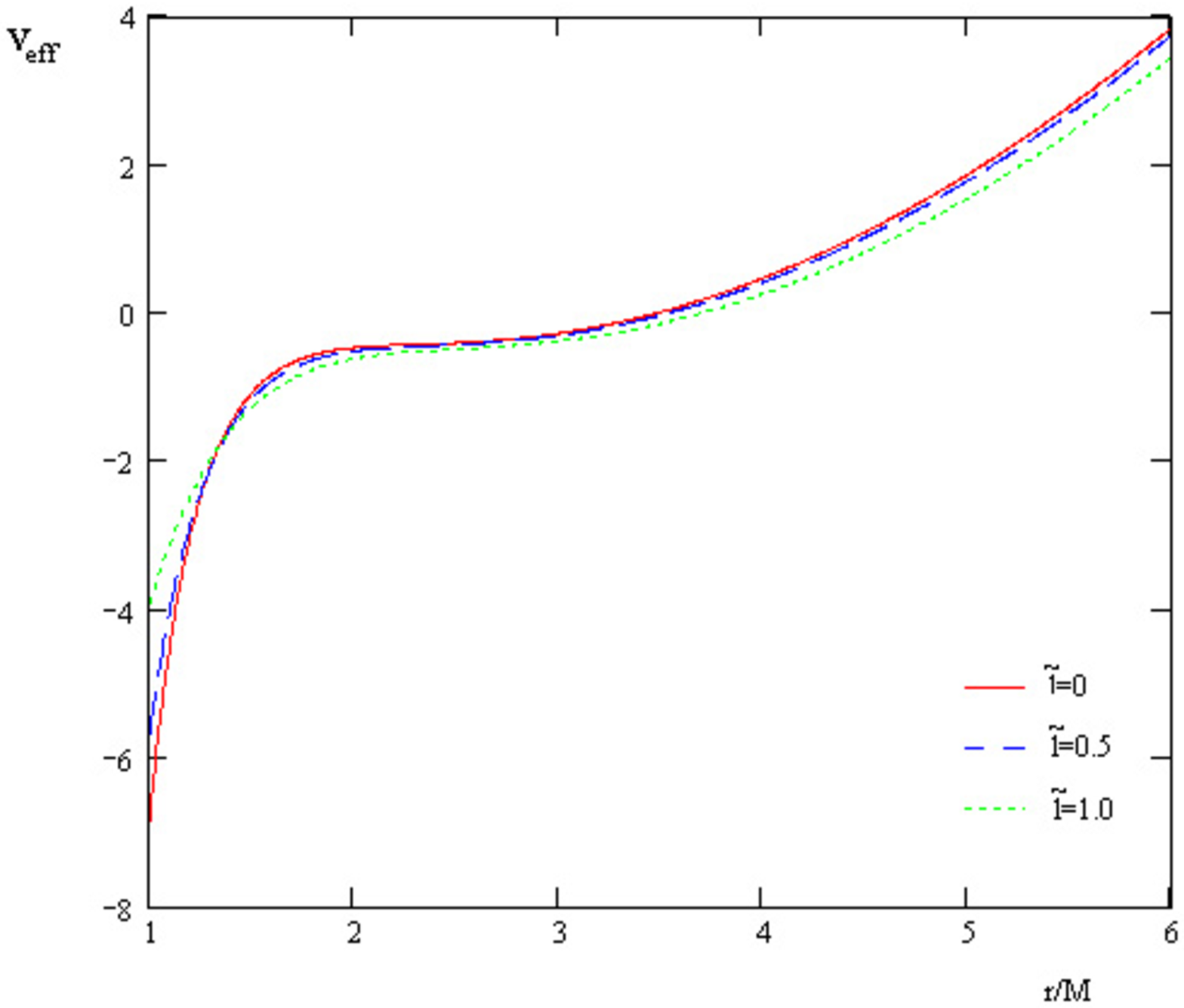}
\caption{{Radial dependence of the effective potential of radial
motion of charged particle near the Kerr-Taub-NUT source in the
presence of current carrying loop around it for the different
values of NUT parameter $\tilde{l}=l/M$ for two cases of modulus
of the dipolar magnetic field a) $\varepsilon=0.3$ and b)
$\varepsilon=0.7$. }}
\label{fig:3}       
\end{figure*}

\begin{table}\label{antilarmor}
\caption{{Stable orbits of test particles near the non-rotating
compact object with gravitomagnetic monopole momentum immersed in
the external electromagnetic field of current carrying loop
(anti-Larmor orbits) }}
\begin{tabular}{llllll}
\hline\noalign{\smallskip}
$\tilde{l}$ & 0 & 0.01 &  0.1 &  0.3 &  0.5   \\
\noalign{\smallskip}\hline\noalign{\smallskip}
$\varepsilon=0.1$ & 3.65296 & 3.65305 & 3.66106 & 3.72475 & 3.84692 \\
$\varepsilon=0.2$ & 3.09353 & 3.09360 & 3.10045 & 3.15506 & 3.26056 \\
$\varepsilon=0.3$ & 2.83420 & 2.83427 & 2.84064 & 2.89154 & 2.99030 \\
$\varepsilon=0.5$ & 2.57592 & 2.57598 & 2.58190 & 2.62931 & 2.72175 \\
$\varepsilon=0.7$ & 2.44347 & 2.44353 & 2.44922 & 2.49488 & 2.58422 \\
\noalign{\smallskip}\hline
\end{tabular}
\end{table}
\begin{table}\label{oddiylarmor}
\caption{{Stable orbits of test particles near the non-rotating
compact object with gravitomagnetic monopole momentum immersed in
the external electromagnetic field of current carrying loop
(Larmor orbits)}}
\begin{tabular}{llllll}
\hline\noalign{\smallskip}
$\tilde{l}$ & 0 & 0.01 &  0.1 &  0.3 &  0.5   \\
\noalign{\smallskip}\hline\noalign{\smallskip}
$\varepsilon=0.1$ & 4.50043 & 4.50058 & 4.51456 & 4.62526 & 4.83569 \\
$\varepsilon=0.2$ & 4.36234 & 4.36248 & 4.37677 & 4.48985 & 4.70439 \\
$\varepsilon=0.3$ & 4.33040 & 4.33054 & 4.34495 & 4.45890 & 4.67491 \\
$\varepsilon=0.5$ & 4.31295 & 4.31310 & 4.32758 & 4.44206 & 4.65896 \\
$\varepsilon=0.7$ & 4.30800 & 4.30815 & 4.32265 & 4.43729 & 4.65445 \\
\noalign{\smallskip}\hline
\end{tabular}
\end{table}

\section{External Electromagnetic Field of Slowly
Rotating NUT Star for a Special Monopolar Magnetic Field
Configuration} \label{star}

 In this Section we will look for stationary
solutions of the Maxwell equation, i.e. for solutions in which we
assume that the magnetic moment of the star does not vary in time
as a result of the infinite conductivity of the stellar interior.
Below we suggest that external electric field is generated by the
magnetic field, taking as a special monopolar configuration. For
this case we can obtain and investigate an analytical solution
with detail consideration of the contributions from the dragging
effects and nonvanishing NUT charge in the magnitude of the
external electric field of the slowly rotating magnetized NUT
star.

 Our main approximation is in the specific form of the
background metric which we choose to be that of a stationary,
axially symmetric system truncated at the first order in the
angular momentum $a$ and in gravitomagnetic monopole moment $l$.
The ``slow rotation metric'' for exterior space-time of a rotating
relativistic star with nonvanishing gravitomagnetic charge is
\begin{eqnarray}
\label{slow_rot} && ds^2 = -N^2 dt^2 + N^{-2}dr^2 + r^2 d\theta
^2+
r^2\sin^2\theta d\varphi ^2 \nonumber\\
&&\hspace{1.4cm} - 2\left[\omega (r) r^2\sin^2\theta +
2lN^2\cos\theta\right] dt d\varphi \ ,
\end{eqnarray}
that is, the Schwarzschild metric plus the Lense-Thirring and
Taub-NUT terms. Parameter $N\ \equiv \left(1-{2
M}/{r}\right)^{1/2}$ is the lapse function,
$\omega(r)\equiv{2J}/{r^3}$ can be interpreted as the
Lense-Thirring angular velocity of a
    free falling (inertial) frame.

As a toy model we could consider the following magnetic field
configuration~\cite{mps01}
\begin{equation} B^{\hat{r}} = B^{\hat{r}}(r) \neq 0 \ ,\qquad
B^{\hat{\theta}} = 0\ .
\end{equation}

Although this form of magnetic field can not be considered as a
realistic, we will show that this toy model can be used to obtain
first estimates of the influence of gravitational field of the NUT
charge on the external electromagnetic field of the star. For this
case, the relevant Maxwell equation reduces to
\begin{equation}
\label{max1a_monopolar}  \left(r^2B^{\hat r}\right)_{,r} = 0 \ .
\end{equation}
The solution admitted by this equation is
\begin{equation}
\label{mf_monopolar} B^{\hat{r}} =\frac{\mu }{r^2}\  ,
\end{equation}
where $\mu $ is integration constant being responsible for the
source of monopolar magnetic field.

Radial magnetic field is continuous at the stellar surface and it
is reasonable to assume that only $\theta$ component of electric
field will survive since it will be produced by cross product of
velocity and magnetic field inside neutron star due to the assumed
infinite conductivity of stellar matter. According to~\cite{ra04}
the interior elecric field is
\begin{equation}
\label{EF_int} E^{\hat\theta}_{in} = -e^{-\Phi}v^{\hat\varphi}
B_{in}^{\hat r}\ ,
\end{equation}
where $v^{\hat\varphi}$ is the velocity of the stellar matter
which is equal to $\Omega r\sin\theta$ for the Newtonian uniformly
rotating star with angular velocity $\Omega$, $g_{00}=-e^{2\Phi}$.

Then the electric field created by monopolar magnetic field is
defined by the following Maxwell equation
%
%

%
\begin{equation}
\label{max1d_monopolar}
 \left(r N E^{\hat\theta}\right)_{,r}
    - \mu {\sin\theta}\left({\omega}\right)_{,r}
    -\frac{2\mu l\cos\theta}{\sin\theta}\left(\frac{N^2}{r^2}\right)_{,r}
=0\ .
\end{equation}

The analytical solution
\begin{equation}
\label{ef_monopolar}
E^{\hat\theta}=\frac{\omega-\Omega}{cN}\frac{\mu}{r}\sin\theta +
\frac{2lN\cos\theta}{r\sin\theta}\frac{\mu}{r^2}
\end{equation}
of equation (\ref{max1d_monopolar}) is responsible for the
electric field of NUT star with the monopolar magnetic field
(\ref{mf_monopolar}). The integration constant
$-\left({\Omega\mu}/{c}\right)\sin\theta$ has been found from the
matching of the exterior solution $\left({C_3}/{rN}\right)$ with
the interior one (\ref{EF_int}) in the Newtonian case taking into
account that the tangential components of electric field and the
radial component of magnetic field are continuous across the
surface of the star. Vector potential being responsibe for the
fields (\ref{mf_monopolar}) and (\ref{ef_monopolar}) is defined as
\begin{equation}
A_\alpha\equiv
    \bigg(0,0,0,-\mu\cos\theta\bigg)\ .
\end{equation}

\begin{figure*}
    \includegraphics[width=0.75\textwidth]{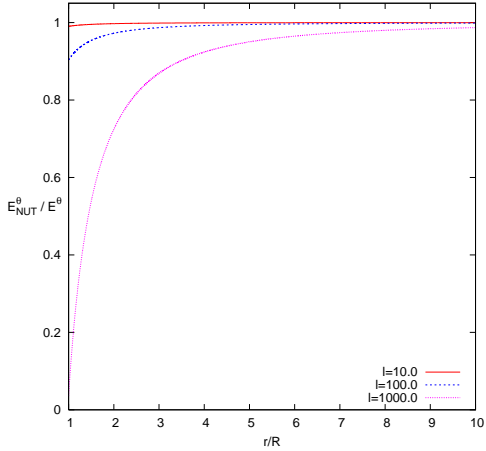}
\caption{Radial dependence of the ratio of the electric field to
that when parameter $l=0$ for the different values of the NUT
parameter. The value of NUT parameter $l$ is given in $cm$.}
\label{fig:4}       
\end{figure*}

In the figure \ref{fig:4} we plot the radial dependence of the
ratio of the electric field to that when parameter $l=0$ for the
different values of the NUT parameter. In this analysis we take
typical parameters for neutron star as radius of star $R=10^6cm$,
$M=2\times 10^5cm$, $\Omega=2\pi/(0.1 s)$, $\omega =
4MR^2\Omega/(5r^3)$, magnetic field at the stellar surface is
$10^{12} G$. Due to the fact that in the right hand side of the
expression (\ref{ef_monopolar}) the first and second terms have
different signs the ratio is less than one. The results show the
strong dependence of the electric field on NUT parameter.

\section{Conclusion}
\label{conclusion}

    Derived exact expressions~(\ref{e1})--(\ref{m2}) for
electromagnetic field in the Kerr-Taub-NUT spacetime indicate that
electromagnetic field will be affected by the gravitomagnetic
charge. However the induced electric field (\ref{e1}), (\ref{e2})
depends on NUT parameter $l$ linearly while the magnetic field
(\ref{m1}), (\ref{m2}) depends on $l$ quadratically.

     {Analytic general relativistic
expressions for the electromagnetic fields external to a
slowly-rotating magnetized neutron star with nonvanishing
gravitomagnetic charge $l$ are presented}.  The star is considered isolated and
in vacuum, and for simplicity with the monopolar magnetic field
directed along the radial coordinate.

We have shown that the general relativistic corrections due to the
dragging of reference frames and gravitomagnetic charge are not
present in the form of the magnetic fields similar to dipolar
case~\cite{go64,ac70} but emerge only in the form of the electric
fields. In particular, we have shown that the frame-dragging and
gravitomagnetic charge provide an additional induced electric
field which is analogous to the one introduced by the rotation of
the star in the flat spacetime limit~\cite{ram01}.

{Motion of charged particles around Kerr-Taub-NUT source immersed
in either (i) external uniform or (ii) dipolar magnetic field have
been investigated using Hamilton-Jacobi equation. We have shown
that in the presence of NUT parameter and magnetic field the shape
of effective potential will be changed. However modifications
caused by external electromagnetic field are dominating.
Investigation of the stability of motion of charged particles
shows, that external magnetic field shifts orbits of test
particles to gravitational source in both cases, while NUT
parameter shifts to gravitational source in the case of uniform
magnetic field and towards loop (in opposite direction) in the
case of the presence of current loop around the gravitational
source.}

\begin{acknowledgements}
Authors gratefully acknowledge Jutta Kunz and Claus L\"{a}mmerzahl
for helpful discussions and comments. Special thanks to Alikram
Aliev for his valuable comments, helpful remarks and interesting
suggestion to calculate the potential differences between the
horizon and infinity. AAA thanks the IUCAA for warm hospitality
during his stay in
 Pune and AS-ICTP for the travel support through BIPTUN program.
This research is supported in part by the UzFFR (projects 5-08 and
29-08) and projects FA-F2-F079, FA-F2-F061 and A13-226 of the
UzAS. This work is partially supported by the ICTP through the
OEA-PRJ-29 project and the Regular Associateship grant. BJA
acknowledges the partial financial support from NATO through the
reintegration grant EAP.RIG.981259. VGK is thankful to DAAD
for the grant A/06/09920 supporting her PhD study in Germany.
\end{acknowledgements}


\end{document}